# Nanostructures in suspended mono- and bilayer epitaxial graphene

Julien Chaste[1*], Amina Saadani[1], Alexandre Jaffre[2], Ali Madouri[1], José Alvarez[2], Debora Pierucci[1], Zeineb Ben Aziza[1], Abdelkarim Ouerghi[1]

[1] Centre de Nanosciences et de Nanotechnologies, CNRS, Univ. Paris-Sud, Universite Paris-Saclay, C2N – Marcoussis

[2] Laboratoire de Génie électrique et électronique de Paris (GeePs), CNRS UMR 8507, CentraleSupélec, Université Paris-Sud, Université Paris-Saclay, Sorbonne Universités - Université Pierre et Marie Curie - Paris 6, 11 rue Joliot-Curie, F-91192 Gif-sur-Yvette Cedex, France

* julien.chaste@c2n.upsaclay.fr 0033(0)649956047

Suspended graphene membrane presents a particular structure with fundamental interests and applications in nanomechanics, thermal transport and optoelectronics. Till now, the commonly used geometries are still quite simple and limited to the microscale. We propose here to overcome this problem by making nanostructures in suspended epitaxial bilayer graphene on a large scale and with a large variety of geometries. We also demonstrate a new hybrid thin film of SiC-graphene with an impressive robustness. Since the mechanics and thermal dissipation of a suspended graphene membrane are strongly related to its own geometry, we have in addition focused on thermal transport and strain engineering experiments. Micro-Raman spectroscopy mapping was successfully performed for various geometries with intrinsic properties measurements at the nanoscale. Our engineering of graphene geometry has permitted to reduce the thermal transport, release and modulate the strain in our structures.

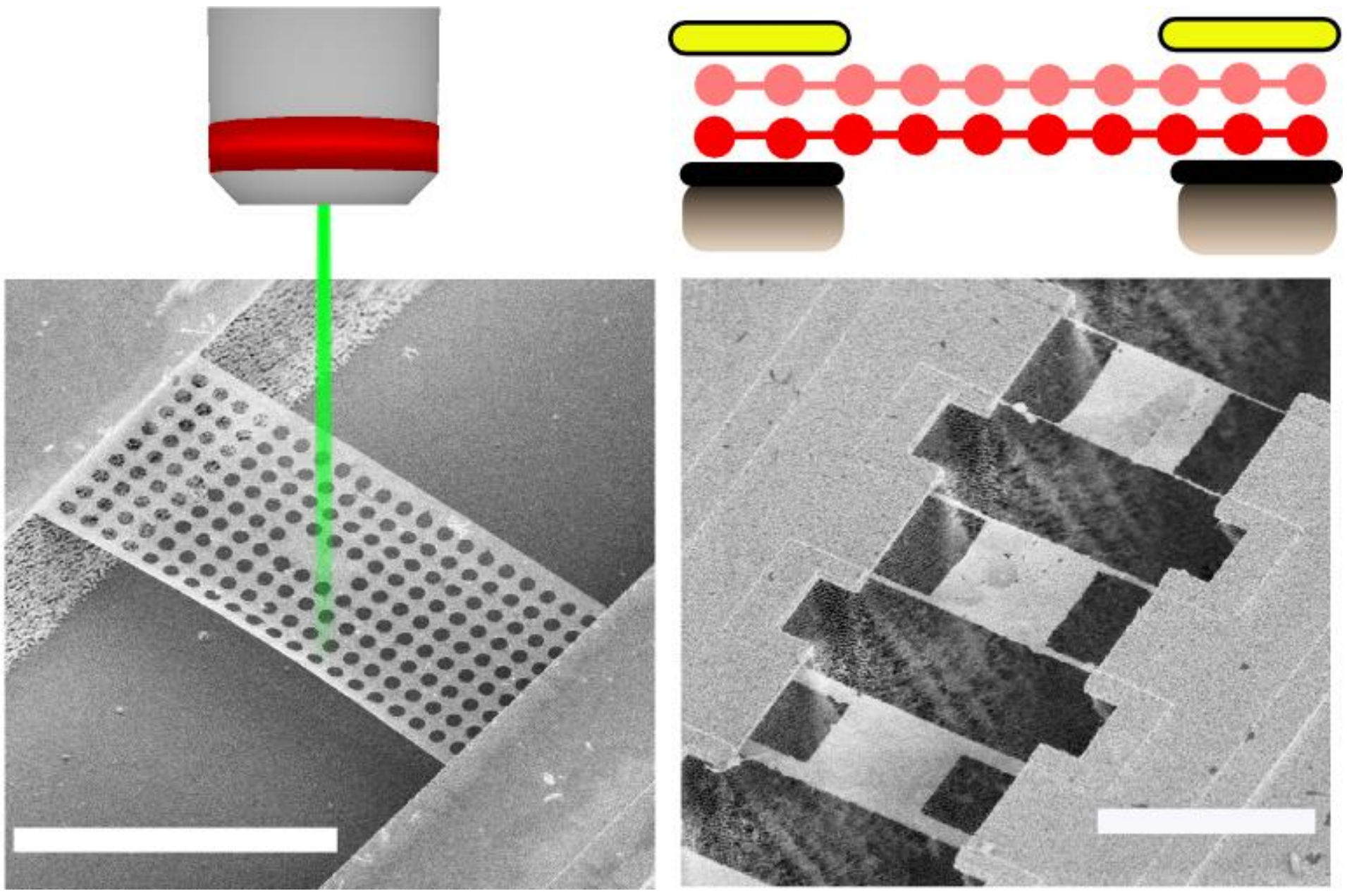

1) **Introduction**

The first realization of suspended graphene was done in 2007 [1,2] , just two years after the first graphene Hall bar samples. It was in fact an incredible feat, and is still today representative of the quality that can be achieved: small monolayer membranes of few microns length which are suspended and resonate with good nanomechanical properties. Until now, the proposed geometries remain quite simple without specific structure in most of the cases. However, nanostructuring of suspended atomically thin material has emerged recently and found various applications: real-time DNA detection, proton, molecular or water filtering, nanoscale kirigami using graphene to obtain stretchable transistors with up to 240% of elongation, a matter-wave beam splitter for molecules trough atomically thin material [3–6]. This can be achieved by lithography and graphene etching, by placing the 2D material on prepatterned micropillar arrays [7,8], or by mechanical buckling of the membrane [9]. There are important motivations to develop nanofabrication techniques of 2D materials. For example, nanostructured graphene with high porosity will permit to combine the thermal transport engineering and the highest thermal conductive material [10]. Nanostructuring can be a straightforward path to control strain gradient at the nanoscale. It is well known to induce strong band structure variation in 2D materials with possibilities to engineer novel systems in electronics and optics[11].

Currently, there is a bottleneck in the design complexity due to low sample numbers, statistical problems, and/or small graphene monodomains. Here we propose an approach allowing fast and simple design of suspended graphene including 2D nanostructuring patterns based on e-beam lithography and epitaxial graphene. Epitaxial graphene offers large scale monodomains graphene and high yield. We have explored the limits of nanostructuring: a quasi-freestanding graphene maintained only by a thin graphene arm. Periodic patterns of nano-opening in graphene or high aspect ratio bars have also been obtained. We show specifically, in cantilever configuration (Figure 2), that we are able to release the native stress existing in epitaxial graphene by 3 orders of magnitude. In order to surpass the limit of structure collapsing in thin film materials, we have created ultrathin film hybrid structures of suspended graphene-SiC (<10 nm thickness) which remain stable, even for very peculiar patterning like zig-zag spring anchoring.

In this paper, we implement not only nanostructures in large graphene membrane but also we modify the intrinsic properties of the membrane like strain and thermal conductivity. We have seen that the physical description of our system is also reduced to nanoscale considerations. We have realized a series of experiments with µ-Raman spectroscopy in order to extract thermal transport, strain and doping properties, and also correlate them at the nanoscale. Raman spectroscopy is largely used for nanomaterial, like graphene, because it is a versatile tool for strain, thermal transport, doping [12,13]. This technique introduces the Stokes or Anti Stokes processes including phonons, photons and electron-holes pair diffusion processes. The well-known G, 2D and D phonons resonances, characteristic of carbon-carbon $sp^2$ binding, the valley degeneracy of K-K' Dirac cones and defects have been investigated here in order to extract graphene properties and notably the strain at the nanoscale.

Moreover, we have managed to measure a reduction of thermal conductance due to nanostructured patterns. This finding will potentially open a large variety of thermal transport experiments in 2D systems. Depending on a specific goal, we are able to control these properties by the membrane geometry itself. Finally, we highlight the possibility to use the asymmetry of both strain and doping existing naturally in the epitaxial bilayer graphene. Such possibility, quite unique in suspended samples, opens new perspective in the tunability and control of graphene nanostructures properties.

## 2) Results and discussion

The methodology of fabrication is similar to the ref [14]. Large scale high quality graphene was grown by annealing the SiC(0001) substrate with a number of layers usually around 1 to 3 [15–18]. Graphene nano-patterning and metallic contact depositions were done by e-beam lithography. The graphene is then released without noticeable damage during the photo-electrochemical (PEC) etching of the substrate and drying. The metal plates serve as etching mask and clamp for the suspended graphene. With this method, we obtained numerous samples with different geometries from 20 nm up to 300 µm (Figure 1 a-f). Standard membranes were bars typically with 6 µm for the length and 3 µm for the width, the success rate to obtain these structures is above 90%. Different opening of holes and rectangle were executed in the structure. We have usually 1500 patterns per millimeter square.

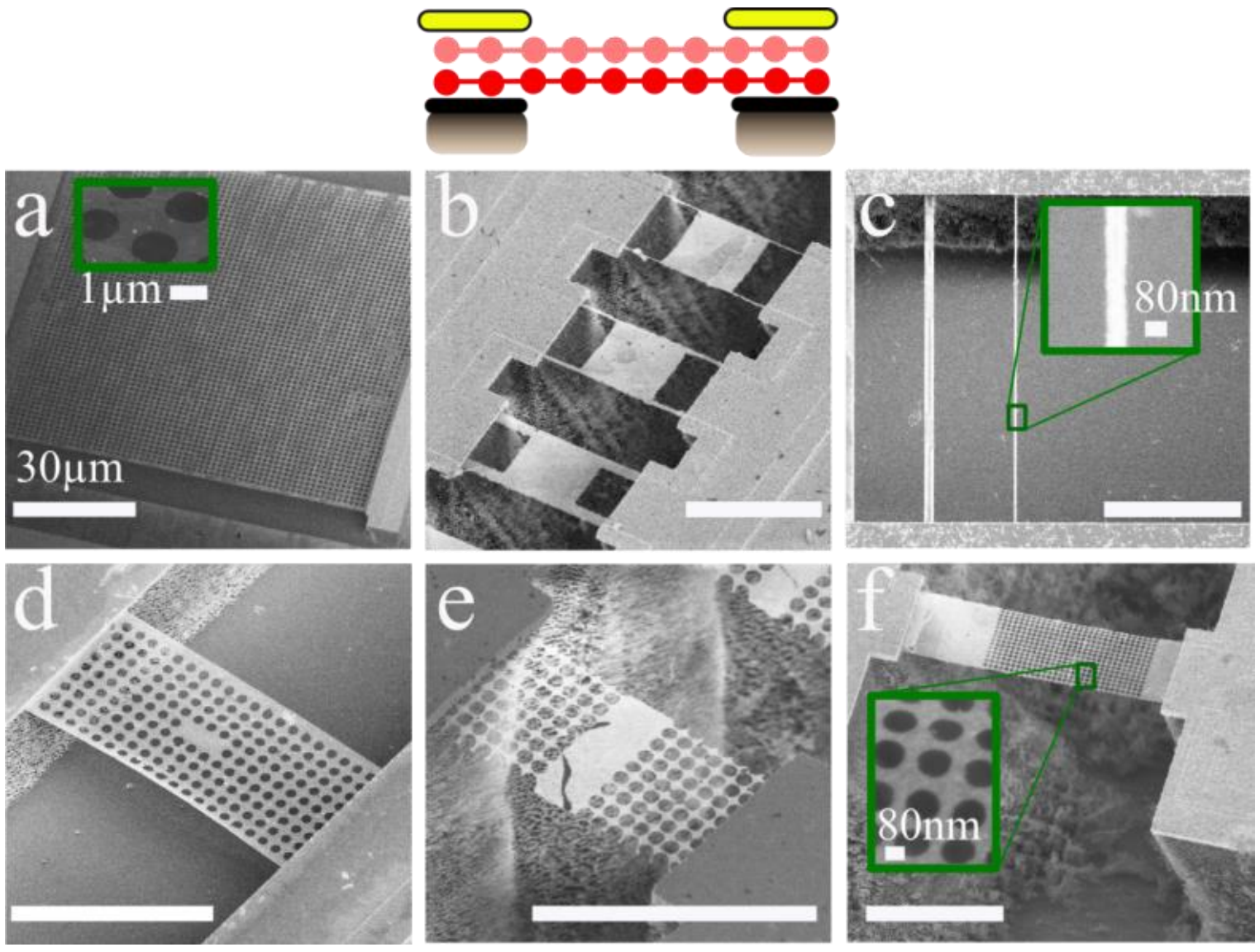

**Figure 1**: **Graphene nanostructuring:** At the top, a schematic of a suspended graphene bilayer clamped at both side with gold and SiC. a-f) SEM images of suspended graphene structures with different geometries from 20nm up to 200µm (at low voltage, 1kVolt, in order to reduce the membrane damage and improve the contrast). **d** represents the possibility of a phononic cavity-like made in a graphene membrane under the Raman laser set-up. **c**,**d**,**e**,**f** are example of holes and thin bar made within clean graphene membrane. The porosity for the "holes area" of structure **e** is around 75%. **a** shows membrane with 100µm side length and 1 µm hole patterns. All scales bar are 5µm unless specified.

We get high geometrical aspect ratio membrane above 100 for the bars of 10 µm length (100 µm), and 80 nm width (500 nm respectively). For holes patterning, porosity exceeding 75% was obtained in a free graphene membrane without any bending. Cantilever with small anchoring area was also obtained. We demonstrated a bended graphene (see SI) but no substantial rolling of the membrane.

The Raman spectra show typically very good quality of the membrane with low D peak (the integrated area ratio between the D and G peak can be above 40), this is better than reported results using similar fabrication methods [19,20]. The ratio of 3-4, between G and 2D peak integrated areas, indicates a rather small and uniform doping; this will be discussed afterwards. The eventual presence of SiC residues is also considered: we are able to etch all the SiC under the graphene in all the samples; but much easier for monolayer sample. For 2 layers or above, we had to calibrate in situ the etching process in order to remove the remaining part of SiC (<10nm) still attached to the graphene without damaging the graphene. This was done by micro-Raman spectroscopy and optical imaging because there is a high optical contrast between a quasi-transparent graphene and a highly reflected SiC film of ultralow thickness (<10nm) (see SI).

In order to go beyond nanostructured graphene, we also investigated hybrid structures which consist in a very thin film of suspended graphene-SiC (<10nm thickness). The nanostructuring of a hybrid plate of SiC-graphene, attached by two zig zag springs from the same material and with high aspect ratio, points out a highly robust film against collapsing, especially when considering the low thickness nature of the system. The structuration illustrated in Figure 2b was specifically chosen in order to highlight the mechanical stability of the hybrid material. We can observe a transparent region in this thin film (delimited by the blue dashed line) which corresponds to suspended graphene. This region shows a lower overall robustness compared to the rest of the SiC-Graphene. These areas correspond to the parallel lines at the SiC steps where the SiC substrate is not completely planar and the dangling bonds of graphene to the substrate are almost absent [21]. In this case, the interaction with the substrate is reduced. This means the complete removal of the thin SiC film is easier in this region.

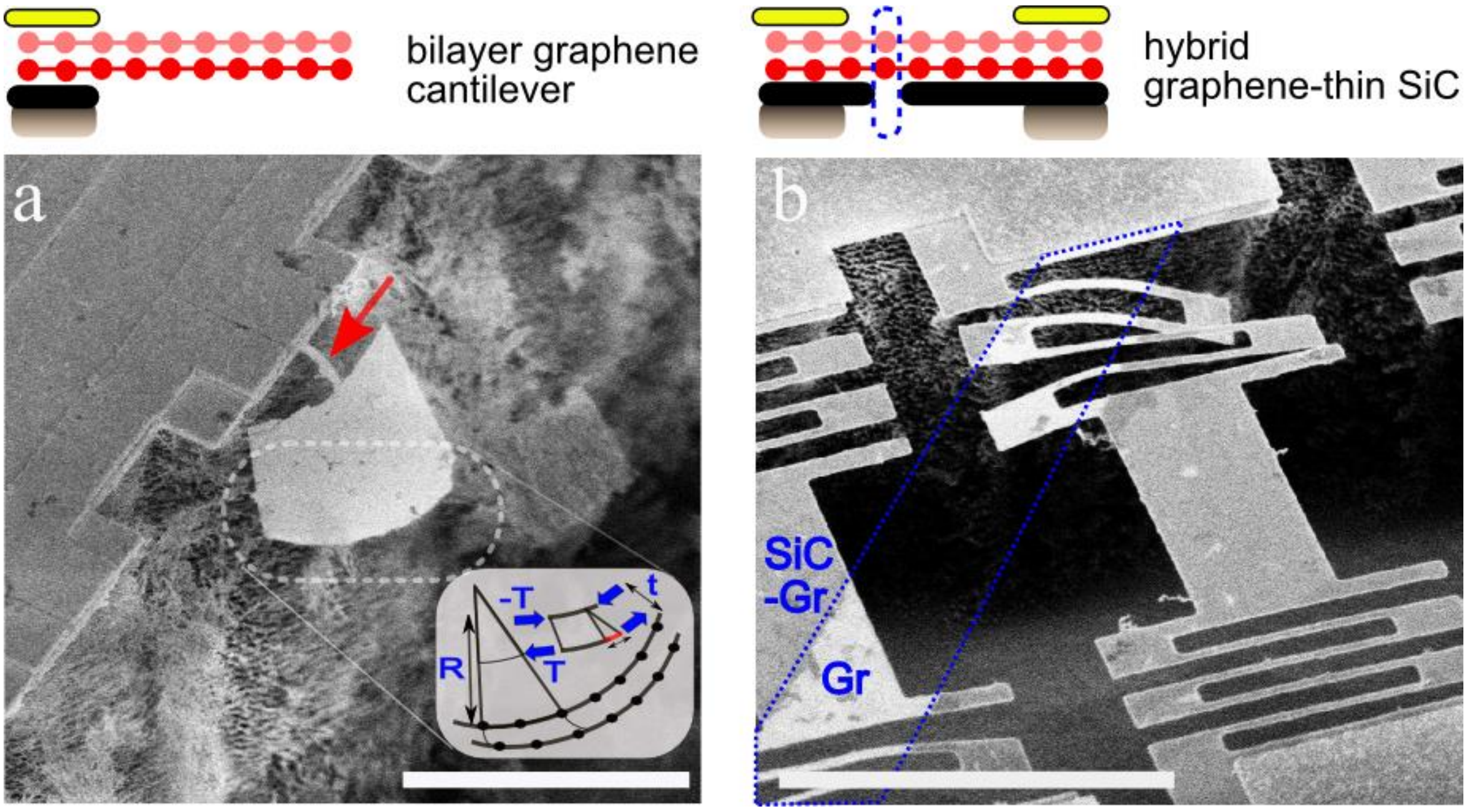

**Figure 2**: **Mechanical stability with extreme nanostructuring. a)**An example of a released membrane at the limit of free-end cantilever shape structure for few layer graphene maintains by a tiny arm (black arrow). In inset a schematic of a curve graphene multilayer, represents the central plate curvature. We show the external layer extension (red portion) due to curvature R and internal shear layer stress T. **b)** Hybrid structures of very thin

SiC-graphene can be realized when the etching of SiC is stopped before the complete release of graphene. These membranes are very robust against nanostructuring as seen in the images with a zig zag spring for anchoring. For example the white region (inside the blue dash line) is graphene only and does not fully sustain the nanostructuring. Scales bar are 5µm.

We focus here on strain properties of our suspended epitaxial graphene. Due to a native mismatching of lattice constants between graphene and hexagonal SiC substrate, the first layer of epitaxial graphene is naturally subjected to a compressive stress of -2.27 GPa [22]. After etching, the gold contacts stay globally fixed. In case of multi clamping, a predefined distributed strain is maintained along all of our suspended graphene structures. This external strain can be largely released in the specific case of a cantilever configuration (Figure 2a). It was possible to obtain unprecedented geometries with quasi-free plates of graphene: few micrometer-dimensions and just attached by a thin arm to the clamping contact in order to mechanically isolate the plate from any external stress. This results in a graphene plate with some curvature. It has to be noticed that most of our graphene flakes have shown much less curvature and in general oriented to the bottom of the sample which mean this specific sample represents the upper limit of internal stress in our cantilever. The residual stress in these specific structures can originate from the interlayer shear stress between graphene layers, the rigidity or stress at the interface may originate from defects within graphene or from residues or external interactions. Uniquely from the curvature R, around 3µm in this specific case, it is possible to estimate the differential strain, $\varepsilon = t/R = 3e\text{-}4$, in a two layers model trough the interlayer thickness t and to quantify a specific stress (see S6). To be noted that the bending due to gravity or electrostatic force can be neglected due to the bending orientation (to the top) the low mass and the low resistive electrical contact between substrate and graphene. Under the approximations of this two layer model, bilayer graphene or graphene on thin-layer-of-defect, we can estimate a residual stress of 100 kPa from an analytical description of the cantilever plate with the Stoney equation[23]. In a perfect device hypothesis, without defect, it is also possible to numerically simulate a graphene flake elongation atomistically when submitted to an interlayer shear stress [24]. If we consider the shear modulus to be 4.6G Pa, the internal stress can be estimated from [24] is around 1MPa. It means that it becomes possible to release the stress inside our sample by at less 3 orders of magnitude with simple nanostructures.

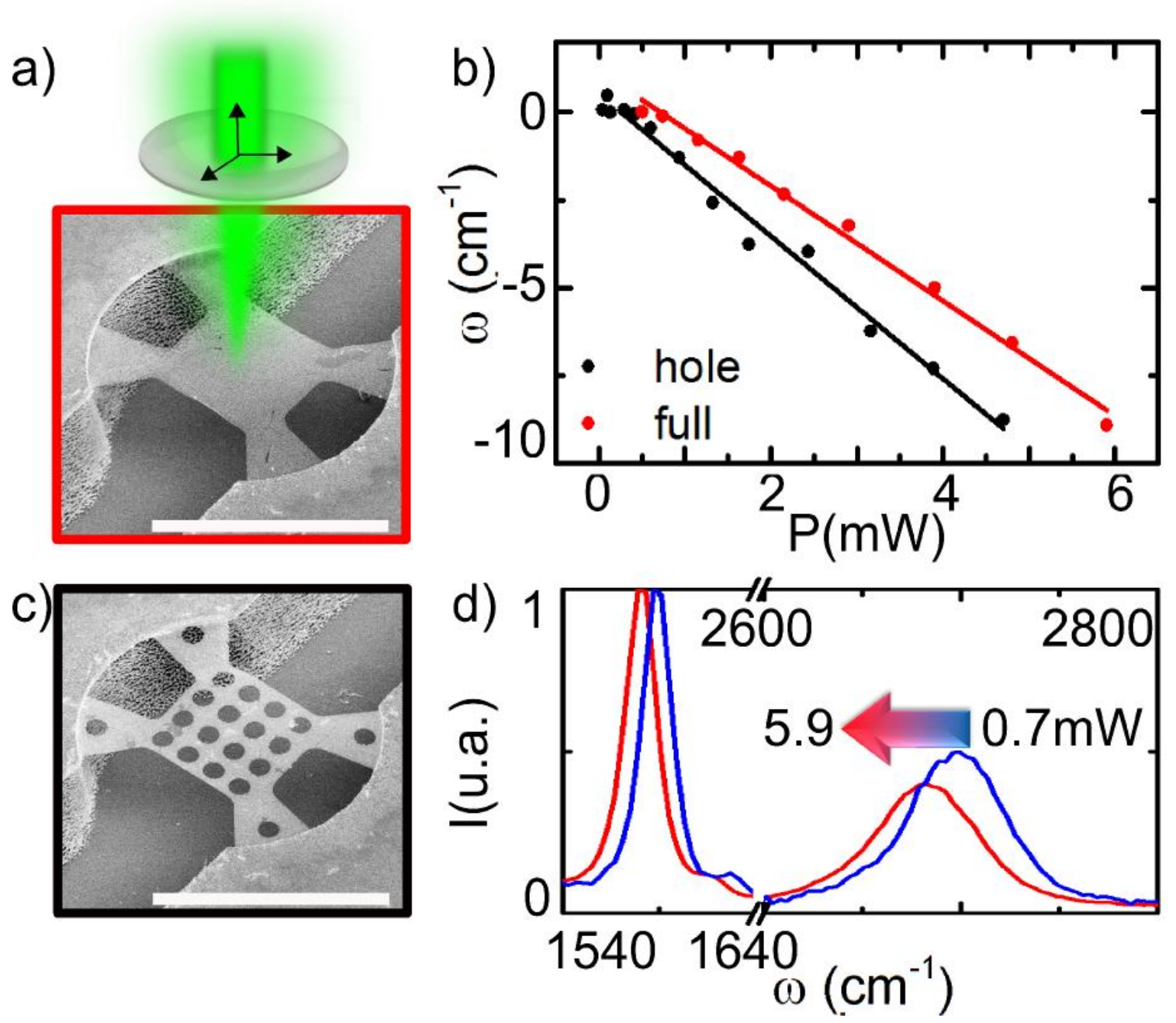

**Figure 3: Laser heating and porosity. a and c)** trampoline shape without and with holes (a and c respectively) Scales bar are 5µm. **b)** Shift of the G peak as function of the incident laser power in mW corrected by the area of interaction ($\nu_2$). The spot was situated in the center of the structure and the slopes are respectively -2.03cm$^{-1}$/mW and -1.6cm$^{-1}$/mW. **d)** Raman spectrum for G and 2D peak as a function of power laser for the hole structure. (additional datas about heating in S4c)

Concerning µ-Raman measurements and graphene nanostructures heating, it is possible to easily extract heating contribution from laser power variation. In fact, this has been used to extract the unusual thermal properties of graphene. Due to the 2D nature of graphene, the thermal transport and phonon behavior is neither diffusive nor ballistic[25] and thermal conductivity unexpectedly depends on the sample size [26]. As a consequence, the geometry of the devices strongly affects the thermal conduction behavior which is still under debate [25–27]. In order to connect the porosity of our nanostructures and their thermal conductivity, we have also studied the Raman peak behavior of our devices under laser heating, at different powers (see Figure 3). The thermal dissipation in graphene has a fast dynamic, with high thermal conductivities[10] but due to the geometrical confinement, it is possible to heat a suspended graphene flake even with perfect thermal contact to thermal reservoir. We did measurements over two monolayer graphene membranes, in a trampoline configuration, the first one is full and the second one has hole patterns over the whole structure. Heating effect on Raman peak position is observed with incident laser power P typically above 1mW at the center of the membrane. This corresponds to an increase of temperature T and thermal conductance k with the Fourier law $P_a=(T-T_0).k$, where $P_a$ is the absorbed power flux, $T_0$ is the ambient temperature. $P_a$ is related to P with a factor $\nu_1.\nu_2$ taking into account the absorbance of the graphene $\nu_1$ and the convolution between the laser spot and the surface $\nu_2$. For the holes, $\nu_2$ is different by a factor 0.81 between the two structures at the same measurement point; which demonstrates the effective reduction of thermal conductance with porosity. For the following measurements, we avoid the strong heating by using a laser power less than 0.1 mW.

Under a small laser power, the µ-Raman peak position is influenced by different properties. A 2D peak position variation of 10 $cm^{-1}$ correspond to around 0.16% of elongation, to a minimum of $10^{13}$ $cm^{-2}$ of doping change and around 500 Kelvin of heating. It is interesting to notice that the 2D peak position is much more sensitive to small stress than usual doping variation. The ratio R of the 2D peak shift and the G peak shift is 0.7 at maximum for doping (it can be even negative for n doping) [28], 1.7 for temperature (it is almost the ratio $\omega_{2D}/\omega_G$) [29] and over 2 for strain (due to additional contribution from Gruneïsen parameters, see Figure 4) [30–32]. This has been used previously for strain and doping separation into Raman signals [33]. We detail specifically the strain with the two main orthogonal directions $\varepsilon_{11}$ and $\varepsilon_{22}$. In literature, only the biaxial and uniaxial cases were measured, but a more general pattern ($\varepsilon_{11}\neq\varepsilon_{22}\neq0$) has to be observed. Nanostructuration is an ideal method to reach this type of strain regime. In fact, $\varepsilon_{22}$ and $\varepsilon_{11}$ could potentially have opposite sign. The hydrostatic strain $\varepsilon_H=\varepsilon_{11}+\varepsilon_{22}$ and its relative $\varepsilon_A=|\varepsilon_{11}-\varepsilon_{22}|$ are the terms which appeared in the Raman signature of the relative G peak position as $\Delta\omega_G=\omega_G.(\gamma.\ \varepsilon_H \pm 0.5.\beta.\varepsilon_A)$. It means two peaks G+ and G- can be observed in uniaxial strain. $\gamma$ and $\beta$ are the Gruneïsen parameter and the shear deformation potential of graphene respectively. $\gamma$ is well established at 1.8 and $\beta$ is around 1 in uniaxial strain case and is related to the anisotropy of the strain tensor. G peak splitting is also important to discriminate the different scenarios as in the following sections.

In order to demonstrate a controlled modulation of the graphene properties by nanostructuring, we have used the Raman spectroscopy to study the electronic properties of the samples. In Figure 4a, a trampoline membrane of monolayer graphene (same growth as in Figure 3) with smaller holes (400 nm) a period of 700 nm has been measured with a laser spot size of 900 nm. In the mapping of 2D peak position and G peak position, we can see a clear pattern which is correlated to the geometry (see SI). There is no noticeable heating effect due to the low laser power and geometry considerations. No visible edge effect was seen in other samples, and the doping peak shift of about 2$cm^{-1}$at the edges observed in [34] do not explain the shift between position 3 and 4 since the edges length are covered by the 900 nm laser spot size at the two points is equal. Thus, we can deduce that this effect is certainly due to strain. To further support this statement, we have done finite element analyses with COMSOL of the hydrostatic strain $\varepsilon_H$ after the release of an initial strain of -2.27 GPa naturally present in epitaxial graphene [22]. The result was smoothed by a 2D convolution with a Gaussian spot representing the laser distribution. The geometry itself was also smoothed with the same method in order to obtain a simulation of the Raman intensity itself. These simulations are in quantitative agreement with our data for the strain modulation due to the holes patterning and the strain in the anchoring of the trampoline. In conclusion, we are able to create a strain modulation inside a nanostructured graphene sample and to confirm it by simple geometrical arguments.

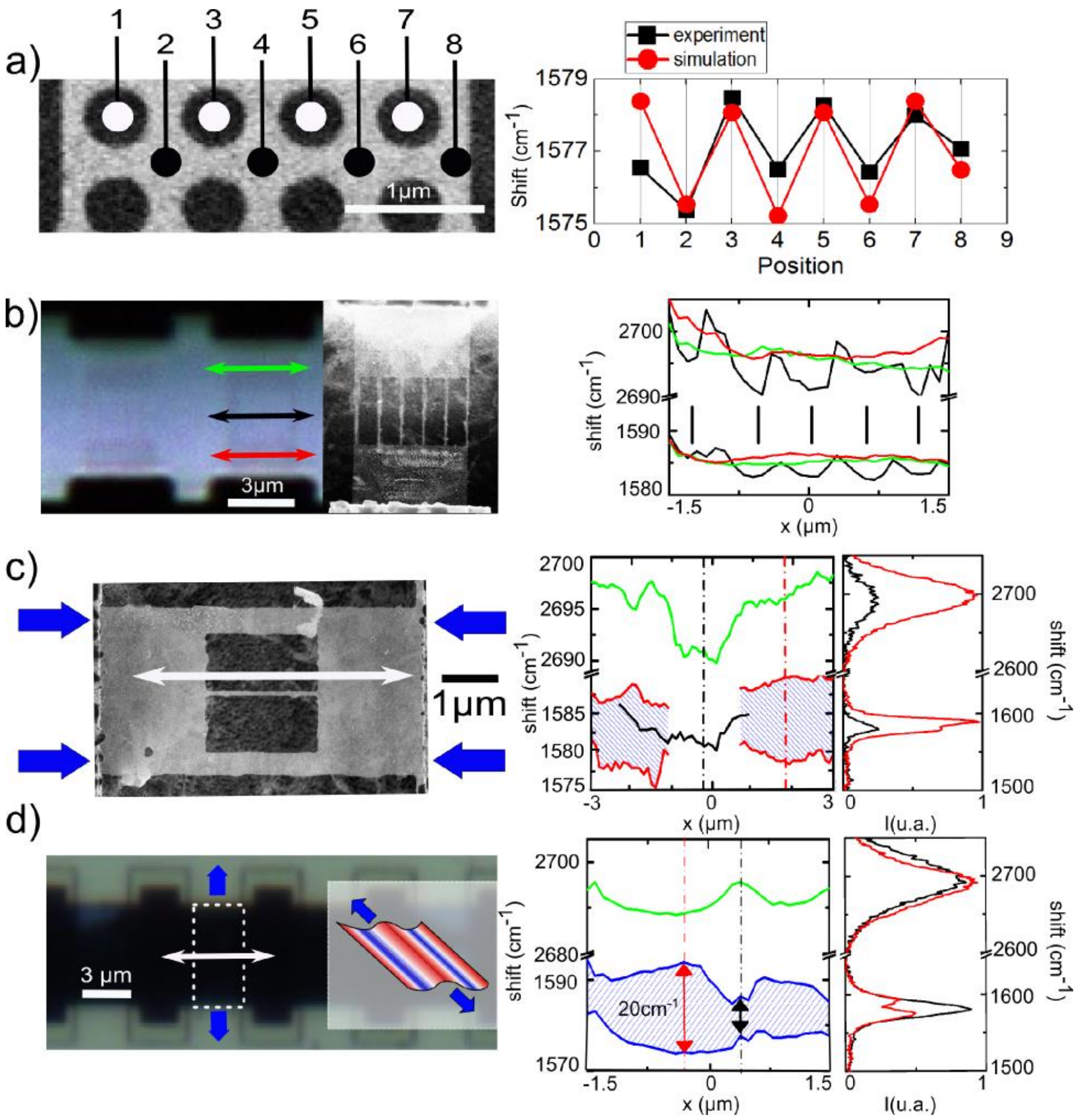

**Figure 4: Four examples of modulation induced by nanostructuring in graphene a)** A similar trampoline membrane as in Fig 3 (monolayer membrane) with smaller holes (P<100µW) and the measure G peak position for the eight positions. A periodic patterns (black) appears and corresponds quantitatively to strain simulation in our membrane (red) (initial stress of -2.27GPa) [22] . **b)** A graphene membrane with a patterning of 6 small bars in the central region. On the right the Raman G and 2D peak position along the red, green and black vectors (only five bars are visible over 6). The modulated signal along the black line is certainly due to strain. **c)** Another structure with a different configuration. Along the white line we see a net modulation of the Raman spectrum between the central part with a quasi-released graphene (black spectrum) and the side part with G peak splitting (red spectrum) (along x we have fit the G peak with one (black) or two (red) Lorentzian curves). **d)** A full bilayer membrane, submitted to strain, probably from a clamp displacement. This type of sample is quite rare: < 1%. On the right, the corresponding Raman spectrum and position along the membrane (G peak is fit with two Lorentzian functions). (P=100µW). Details and additional datas are in the SI

The second example represents a periodic nanostructuring of small bars created in a rectangle membrane of a bilayer graphene (Figure 4b). From now on, the used laser spot size is around 500 nm. In the bar region, small shift are observed for the 2D and G peaks without splitting. The ratio R is around 2-2.3 which corresponds to a relative strain measurement. The relative 2D peak shift of -6.6 $cm^{-1}$ corresponds to a relative strain $\varepsilon_H$ of +0.1%. Two scenarios are possible; the central bars are stressed (released) and the full regions are released (compressed). In this work, additional datas where obtained for few similar membranes with similar results. We tend to assimilate this small positive strain to the top layer of our bilayer graphene with natural strain asymmetry. This layer is initially without noticeable stress but is stuck by van der Waals interaction to the bottom one, which is compressed by -0.2%. We assume when the compression inside the bottom layer is released, the energy is transferred to curvature or buckling of the whole membrane and also in a smaller and opposite stress in the top layer. It is common for epitaxial graphene to have a less effective signal coming from the bottom layer than the top layer [22] (in monolayer) (see SI).

In the last examples of Figure 4c and 4d, we show Raman spectra modulation with complex features. This implies a strong G peak splitting coming from anisotropic planar strain ($\varepsilon_A \neq 0$) [30–32] or asymmetric doping [35–37,20] (for 2n layers graphene) or asymmetric strain between layers (see also S10 for additional measurements and a complete analyze). Figure 4c represents a scan along the rectangle membrane in y-axis. It is similar to Figure 4b with two larger bars on each side and one thinner in the middle of width W. A large splitting is observed only in the middle of the two full parts like in 4 other similar samples with small W difference (see SI). Our example is considered as the most favorable for a high G peak splitting. In Figure 4d, we present a full membrane where strong modulations of the G peak splitting and peak shift are observed. This sample is statistically quite rare, 90% and more of our full membranes have no modulation and no splitting at all. This type of modulation has been seen before in wrinkled graphene with the formation of an orthogonal buckling wave [9,38]. This means that we have certainly created an artificial strain on this sample due to a small and rare movement of the contact anchoring.

For both samples, doping variation alone does not explain the result since it would mainly represent a huge doping asymmetry of few $10^{13}$ $cm^{-2}$ and an unexplained strong correlation with the graphene geometry. Moreover, simple planar strain simulation of a monolayer membrane without buckling didn't match with the measured datas. In the case of full buckling, the stress energy must be quasi-released and the strain-induced peak shift and splitting must disappear. Therefore, we explain our results in a complex mixing of two effects: 1) the bilayer graphene with asymmetric strain and doping on each layer. A small buckling of the strained bottom layer induces a curvature of the stick top layer and an opposite strain. This can be effective especially when wrinkles are appearing. 2) In plane strain orthogonal components which, in our devices, can be of opposite sign. Strain matrices in each layer present a general planar anisotropy for each point where orthogonal strain can be of opposite sign. Further information is needed here to analyze the datas like the 2D' peak shift at 3250 $cm^{-1}$ or polarized Raman measurements. We can also imagine simpler test structures which are compatible with this type of analyses like in Figure 4a and 4b.

## **3) Conclusion**

In summary, we have successfully demonstrated different possibilities of suspended graphene in order to create atypical structures with new geometries. We show the effect of graphene nanostructuring with different patterns on thermal conductivity reduction and strain modulation using μ-Raman spectroscopy. These results can be used to achieve nanostructures with other 2D materials for example for large monolayer of $MoS_2$ or large monodomains of CVD graphene bilayer. This study about nanoscale strain modulation in suspended 2D membranes is a key issue in nanomechanics and optoelectronics. Particularly, it combines perfect mechanical materials with strong mechanical coupling, high external sensitivity and thermal dissipation tailoring. Beyond this work, the development of new type of graphene nanostructures would allow to probe the reduction of thermal conductivity by few orders of magnitude. This can pave the way to understand fundamental properties such as the abnormal phonons-phonons scattering mechanism in 2D materials with nanostructuring or even specific phonons specularity on nanostructured edges[39]. We can also use it to create concentration of strain in a nanobridge with the 2D equivalent of an anvil cell.

**Acknowledgement**

This work was supported by ANR HD2H grants, as well as a public grant overseen by the French National Research Agency (ANR)

# Supplementary information

## S1: Fabrication and other examples of nanostructuring

**Fabrication of nanostructured suspended graphene**

The graphene is epitaxially growth on the Si face of a high quality n doped SiC ($2.10^{17}cm^{-2}<n<2.10^{18}$) at 1500°C in a furnace. A hydrogenation step was usually added afterwards at 800°C for 10 min under argon and hydrogen to decouple the graphene layer from the substrate. Alignment marks and pre-contacts patterns in Cr/Au were deposited by e-beam lithography technic with PMMA directly on SiC. Graphene was previously remove underneath by 40s $O_2$ RIE plasma. The graphene mesa and the holes were etched in graphene afterwards, again with $O_2$ plasma. Clamped contact patterns were deposited at the end with Cr/Au over graphene and pre-contacts. All the graphene structures were released in a KOH bath (0.8%wt) during a photoelectrochemical (PEC) etching of 2-8 µm of SiC at ambient temperature, under UV insolation and current around $1mA/cm^2$. Etching rate was 1-1.5µm/hour. Some additional step of etching was added afterwards, with sample still in liquid and improving current, until the SiC was completely removed from the graphene (see SI section S2). This was observed, after a certain etching time, under an optical microscope and additional step of etching were done if necessary. We measured the UV power to be $1.5mW/cm^2$ on the sample. The final drying was done with a critical point dryer.

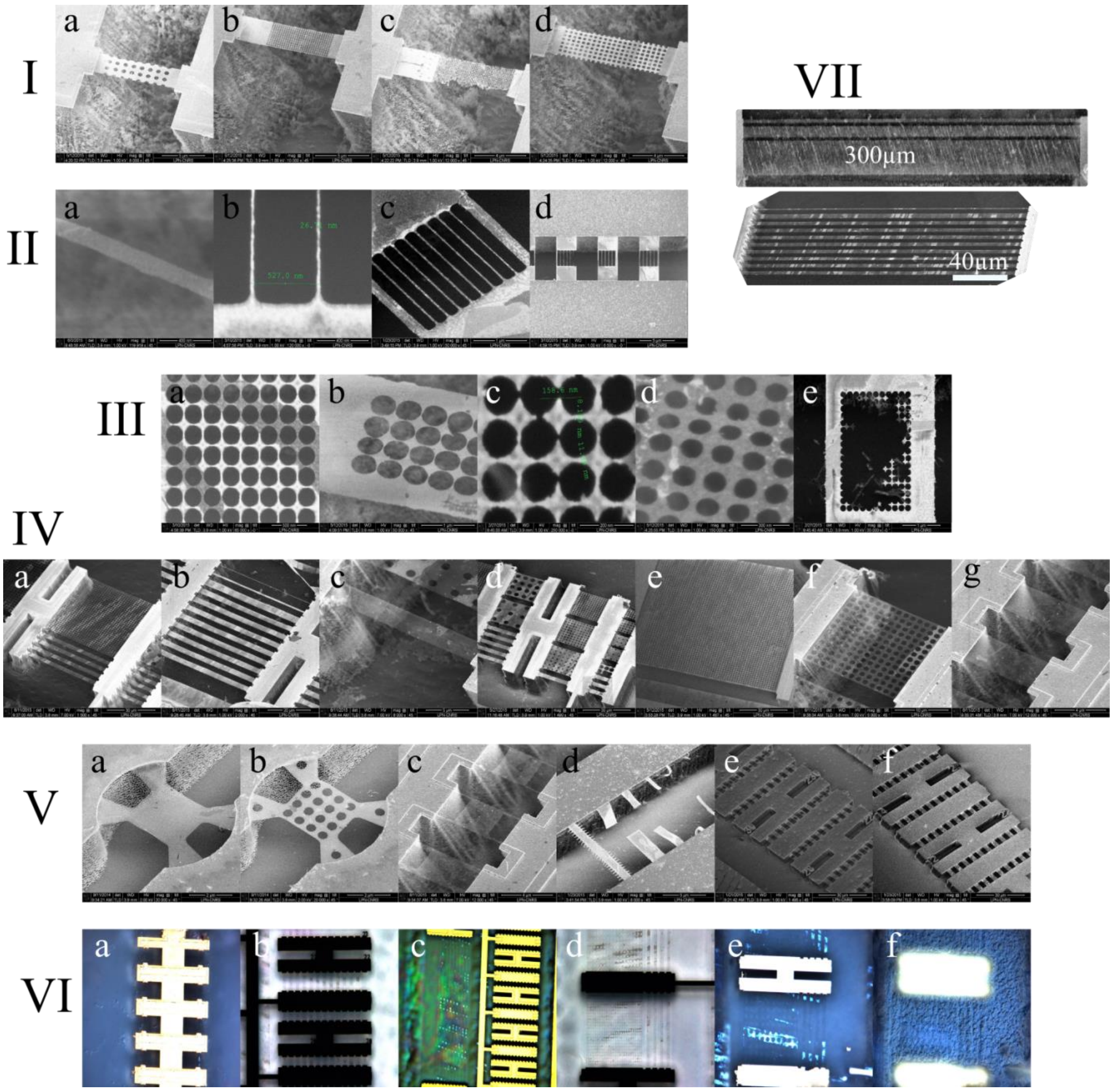

**Figure S1:** SEM (line I to VI) and optical (line VII) images of diverse suspended graphene samples. SiC was completely removed for small structures (< 10µm) and quasi removed for very big structures. I) bars of 10µm by 3µm with different holes patterning. II) Some bars of 6 by 3µm with 2µm length suspended bar pattern in the middle part. III). Test of maximum porosity in suspended graphene exceeding 75%, in c and e we can see the limit for the size of anchoring between holes IV) Patterns in rectangle or squares with or without holes from 10µm to 100µm side (e). V) (a,b,c) Complex structuring with a trampoline shape (a middle part maintain by few thin branches) (d) Cantilever. (e,f) Some of the typical 6 by 3µm bar with nanopatterning in the middle VI) Optical images of some samples in order to proof the negligible presence of SiC (see section S2-S3).

## S2: Complex structuring of a new thin hybrid graphene-SiC

On contrary to a complete etching as in fig S1 or ref [1] and other references [2–6] on the PEC technic, with a simple step of etching (one current around $1mA/cm^2$), most of our observation have shown the presence of a thin layer of SiC (5-15nm) under the graphene. It is interesting to notice that the result is quite rigid and homogenous, even with this small thickness and complex structuring was already achieved (see figure S2). This suspended hybrid structure between graphene and SiC can represent a real opportunity to improved geometrical aspect ratios in thin membranes of SiN or SiC and is quite interesting by itself.

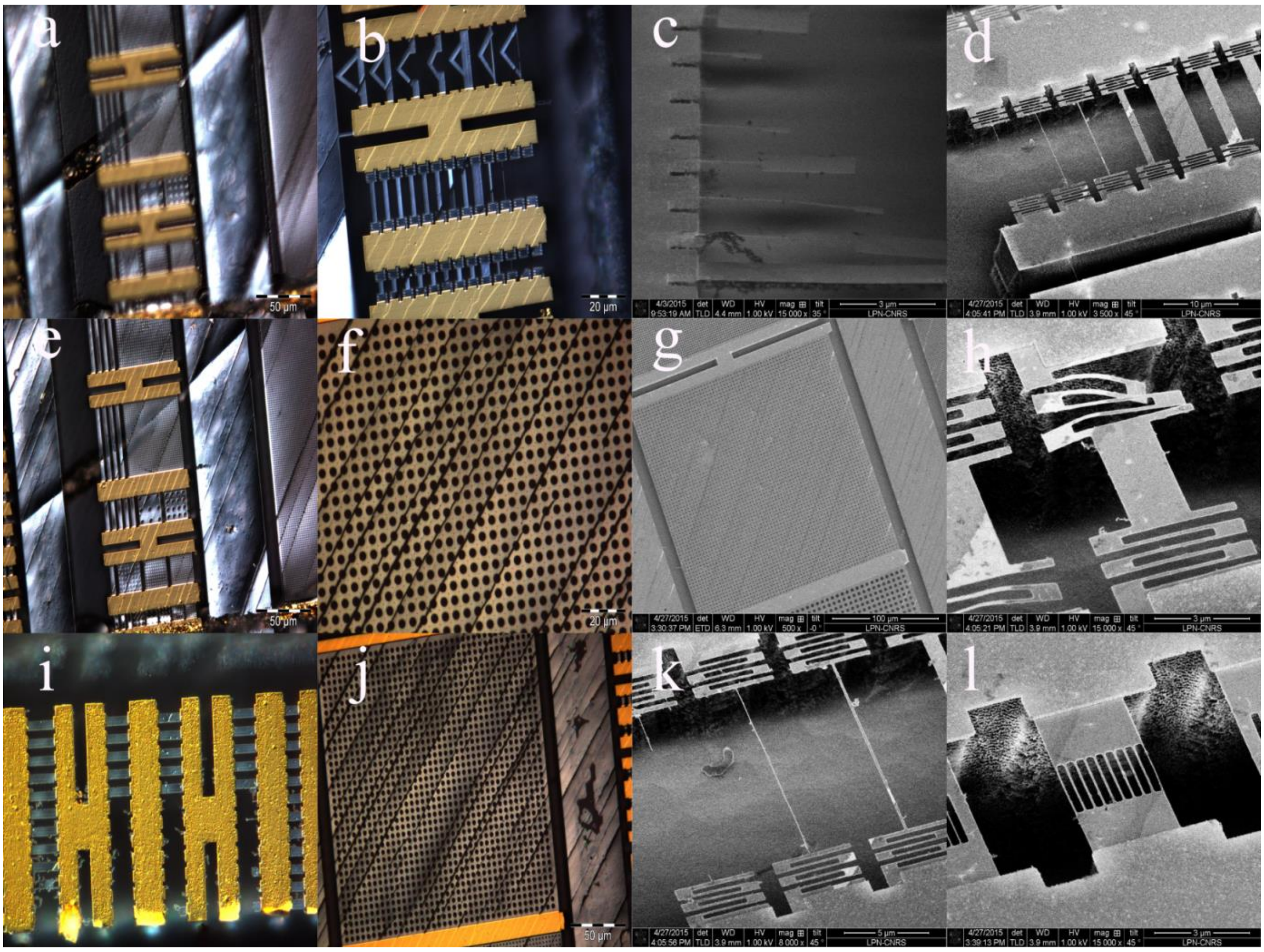

**Figure S2 :** SEM (c,d,g,h,k,l) and optical (a,b,e,f,i,j) images of diverse suspended graphene samples. It is possible to achieve square of 200µm by side (f,j), the contrasted cross-line correspond to n and n+1 layers of graphene and demonstrate that the thin SiC layer is easier to remove on the n+1 layers. (b,c,i) cantilever with V-shape of simple bar, in c we can see the deflection of the cantilever. (b,d,h,k) structures with a center part maintains by zig-zag anchoring arms in order to uncoupled and completely released the center part from external tension, especially anchoring dissipation and stress.

## S3: Etching control and optimal conditions

In order to remove the thin SiC layer efficiently without modifying the graphene layer, we have add some steps during etching with increasing current until from 1 to 5mA/cm$^2$ . The time and current of PEC etching was determined for each sample with a direct observation under an optical microscope (between step and with sample still in water) (figure s4 is about the determination of this SiC film thickness and its calibration with an optical microscope). Typically, a 6-8μm etching of SiC and complete removal of SiC over small structures and optimal conditions was obtain with step of 30mins and improving current of 1mA/cm$^2$, 2mA/cm$^2$, 3mA/cm$^2$, 4mA/cm$^2$, sometimes 4.5-5mA/cm$^2$. If the current increase is too high, the graphene can be damaged like in fig S3c. A reasonable compromise was found, which consist to stop the etching when most of the SiC is removed from the graphene but not at 100%. This is the reason why big structures still have some SiC traces. With this methodology, for small structures, homogenous structures without SiC trace were obtained.

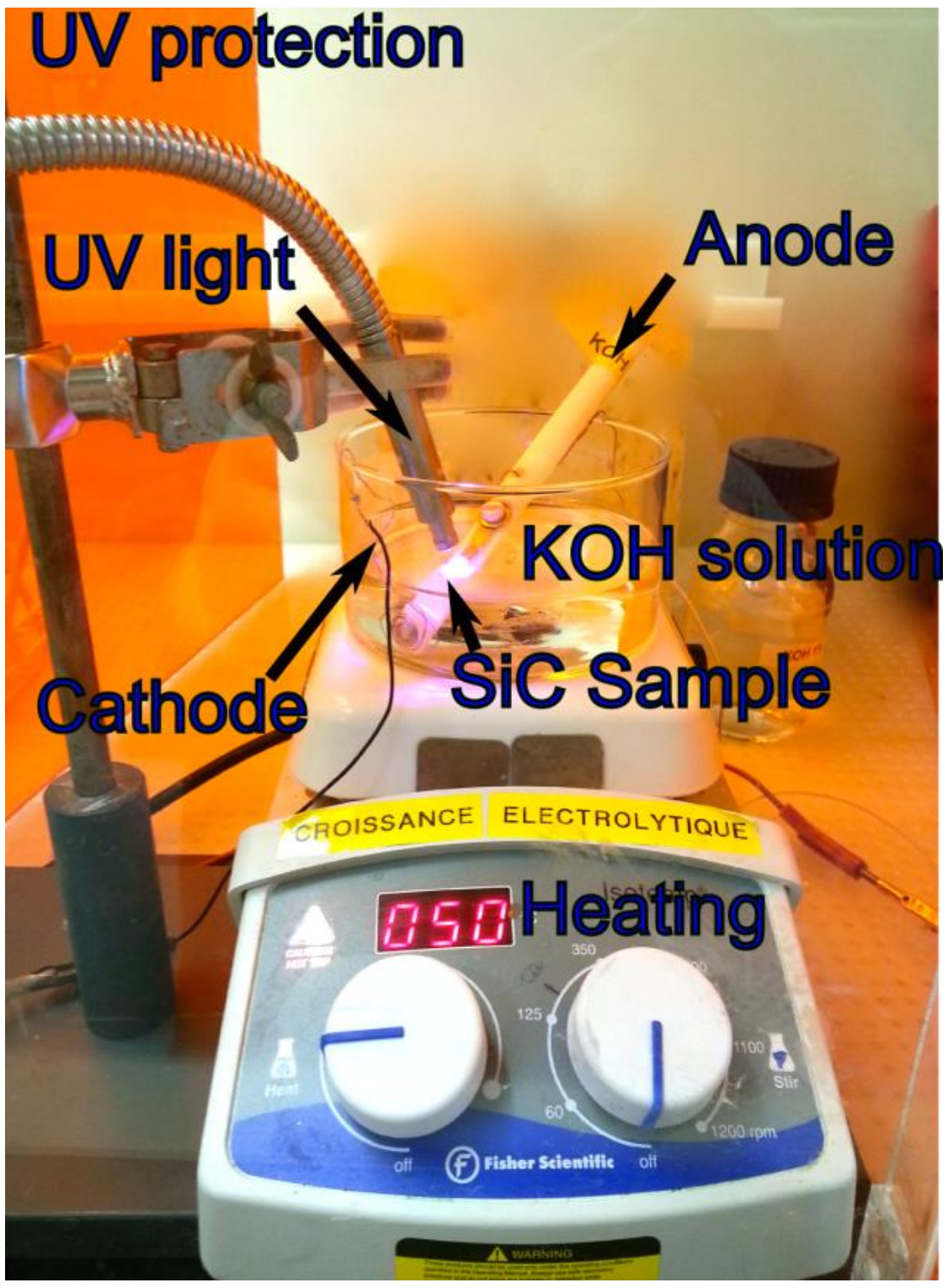

**Figure S3a:** PEC set-up for SiC etching under UV light, electrolyze potential configuration and within a KOH solution bath

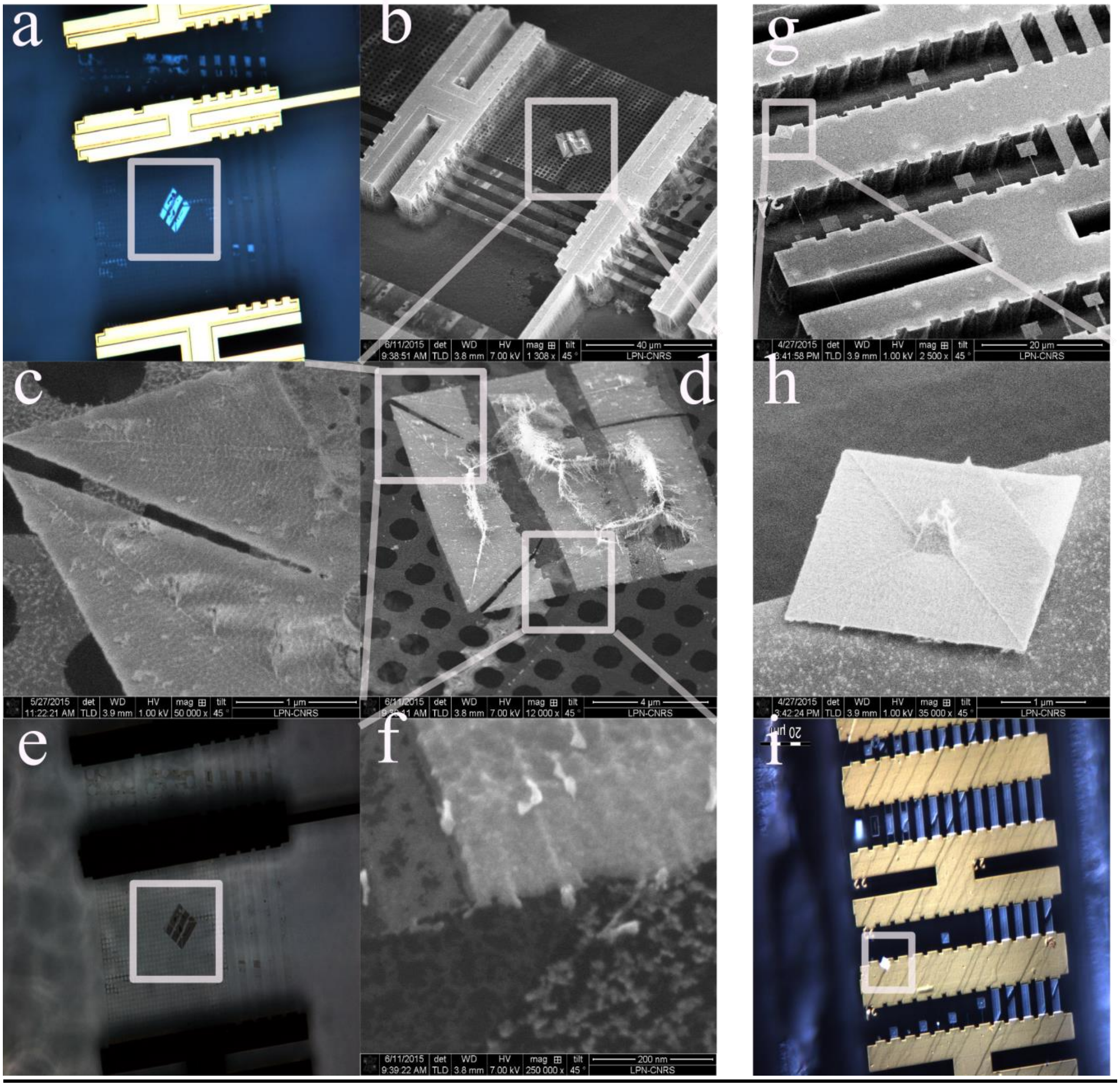

**Figure S3b:** SEM and optical images of suspended graphene. **a-f:** a graphene-SiC hybrid membrane has been flip and deposited by chance on top of a clean graphene membrane of 50µm side (**b**). It is possible to see the graphene, the residual SiC and the SiC filaments with SEM (**d**). The SiC thickness is around 10nm (**f**). The membrane is quite clean by comparison with a focalized image at the sample stage (**a**) and the substrate surface stage (**e**): the membrane is clearly quasi transparent, on contrary to nanometers SiC residuals. Although some part of nanometer SiC parts still stick on it (black arrow in **a**). **g-i:** a cantilever was flip on gold patterns. It is possible here to check the thickness of SiC for the hybrid graphene SiC structure (~10nm) and its good quality. A SiC base appeared in the middle and a corner was released free of SiC (black arrow in **g**).

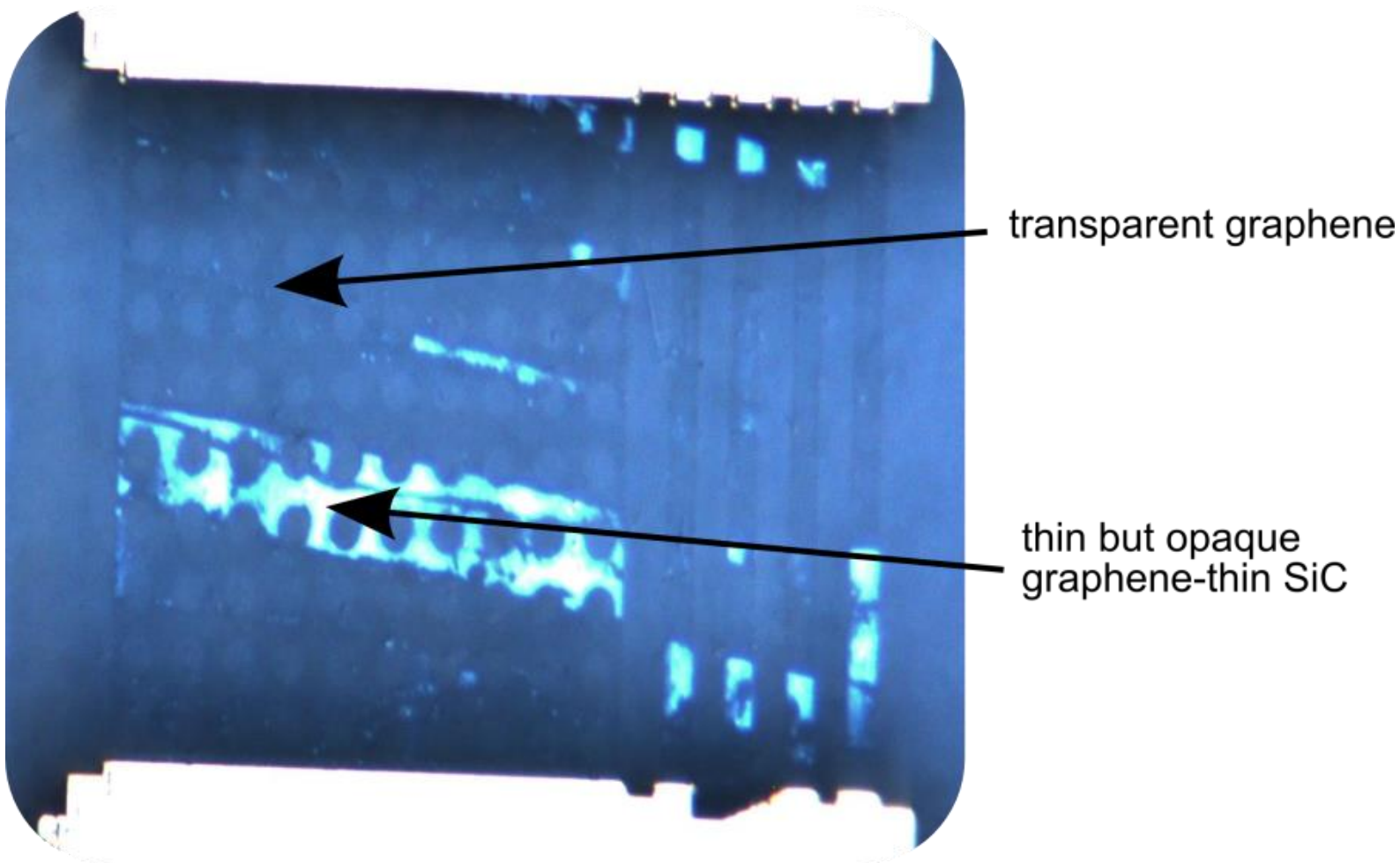

**Figure S3c:** Optical image of several graphene membrane with holes in order to show the contrast between the transparent area of graphene and opaque buth thin (<10nm) of graphene-SiC hybrid structure.

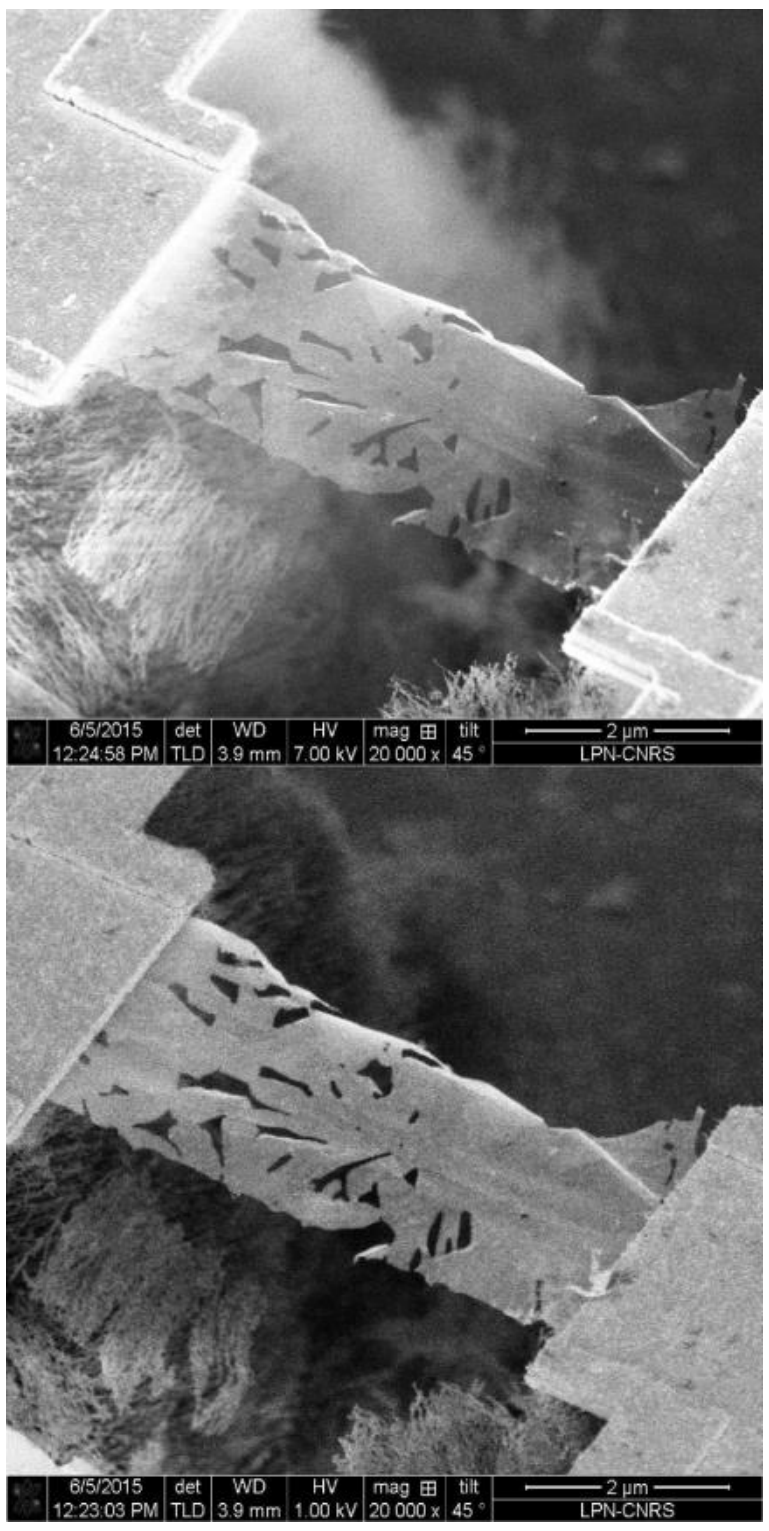

**Figure S3d:** SEM of a suspended graphene. When the PEC etching is somehow too efficient, with too much current for example during the etching, the graphene begins to break and some holes are appearing in the graphene membrane.

## S4: Overview of our graphene; layers number, quality, quantityof the substrate

Raman spectroscopy has been used intensively with graphene and carbon nanotube. It is a technic sensible to many physical aspects of the measured material, especially carbon-based ones. The relative intensities, FWHM and positions of usual graphene D, G or 2D peaks will change depending on the number of layers, stress, doping, and temperature or defects density like graphene edges… These resonances can be seen in an example of a Raman spectrum of our samples in figure 1 in the main text.

**SiC contribution:** In addition, SiC peaks contribution is the bulk substrate signature few µm away from the graphene and these peaks analyses combine with the optical and electronic images have shown no evidence of SiC stick to the bottom of the graphene layers (**see S4b**), unless specified in this paper. When the confocal Raman set-ups are well focalized at the graphene level, because the membrane is fully suspended in our case, we have a high sensitivity of typical peaks level and also observe small N and D* peak at 1740, and 2400$cm^{-1}$ with a very low contribution of SiC substrate compare to epitaxial graphene on substrate. The 3 peaks, around 700 $cm^{-1}$, are main signature of the bulk substrate. It usually varies at less than 2% out or inside the lateral graphene position. In a similar manner, when the laser is focalize on the bottom surface of SiC, the signal of graphene peaks is close to the noise floor and change also by 2%. We interpret this by an optical resonance in the cavity between graphene and the surface due to the absorption of graphene which is 2.7% by graphene layers. It means we don't see any significant signature of SiC onto graphene in most of the case present here unless specified. It has to be add that the optical analyses of part S3 corroborate this analyze.

**First and second layer intensity ratio:** In our measurements, the bilayer graphene emits 3.5 times more than the one layer graphene in Raman signal intensity. This large value is reasonable considering other publications on epitaxial graphene [3,6–8] where the 2D peak is, at less, doubled in bilayer graphene and this can be attributed to the interaction with the residual buffer layer [10] which change doping and photon emission. This interaction is stronger on the first bottom layer than on the second layer due to electrostatic screening. We assume it explains why we have to systematically introduce a factor 2 to 3 for the intensity in the bottom layer and in the top layer in our simulations of Raman measurements in asymmetrically strain bilayer graphene

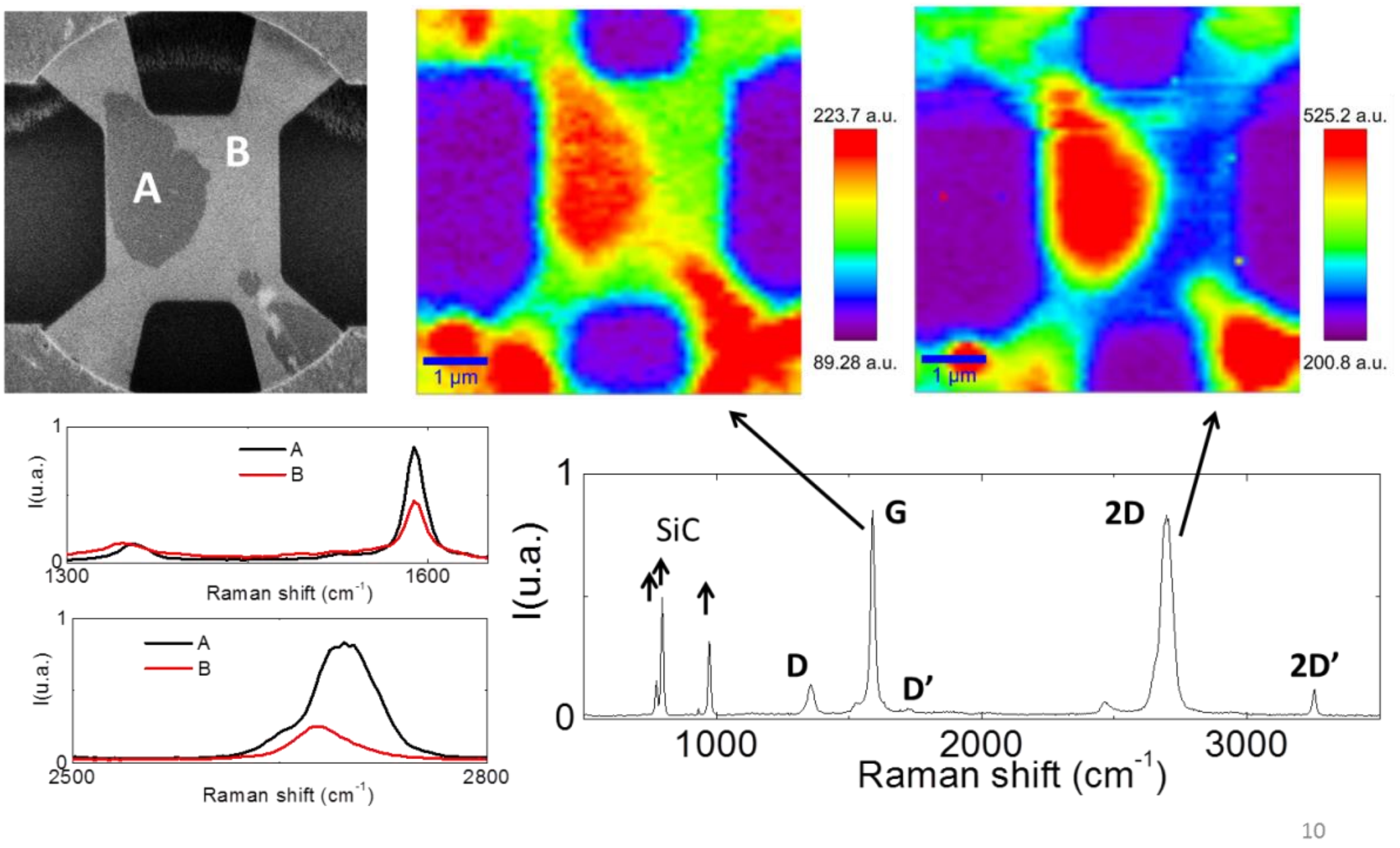

**Figure S4a:** SEM picture of a trampoline graphene membrane with a part were a graphene layer was removed. Monolayer (in red) an bilayer (in black) graphene present a different shape of the 2D peak: a perfect Lorentzian shape for the monolayer (see ref [11–13]), and an additional contribution on the left part of the main peak for the bilayer. It can also been seen

in the intensity of the Raman peak (G and 2D). The Part B emitted 3.5 times more than the A part concerning the peak 2D intensity. In the Raman spectrum, we can see contribution from different peaks characteristic from graphene and especially multilayer graphene and also SiC substrate. The difference between the part A and B is also shown in a zoom for the D, G and 2D peak.

**<u>D-peak and defect:</u>** In our suspended graphene membranes, we can see the D peak is low or not even apparent with a ratio of 40 in intensity if we compare to the G peak integrated intensity.

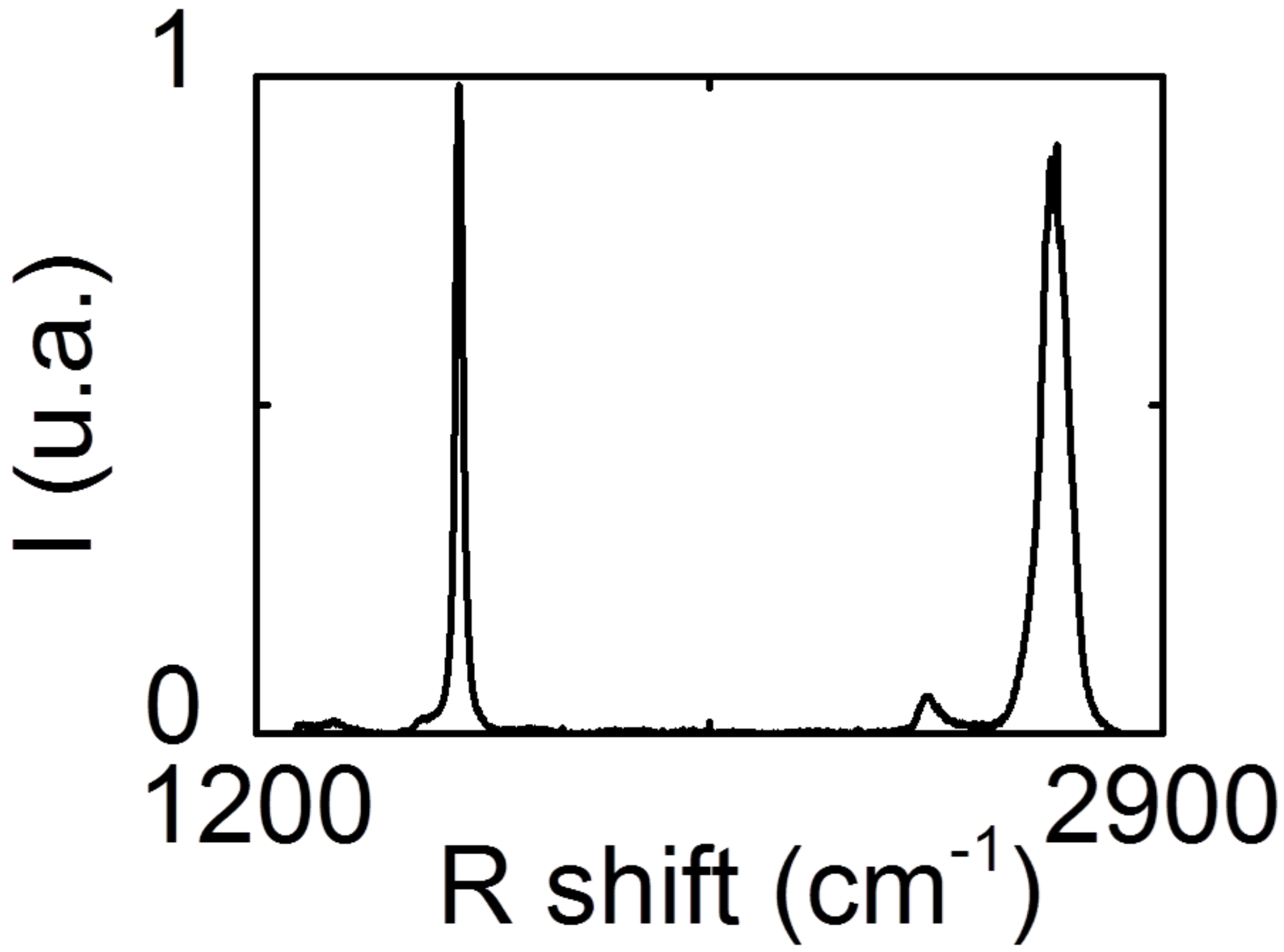

**Figure S4b:** Raman spectrum example with a ration ratio between G peak and D peak around 40. In average on our samples, this ratio is around 7. The value are mainly around 30 expect certain zones with a ratio averaging of 7.

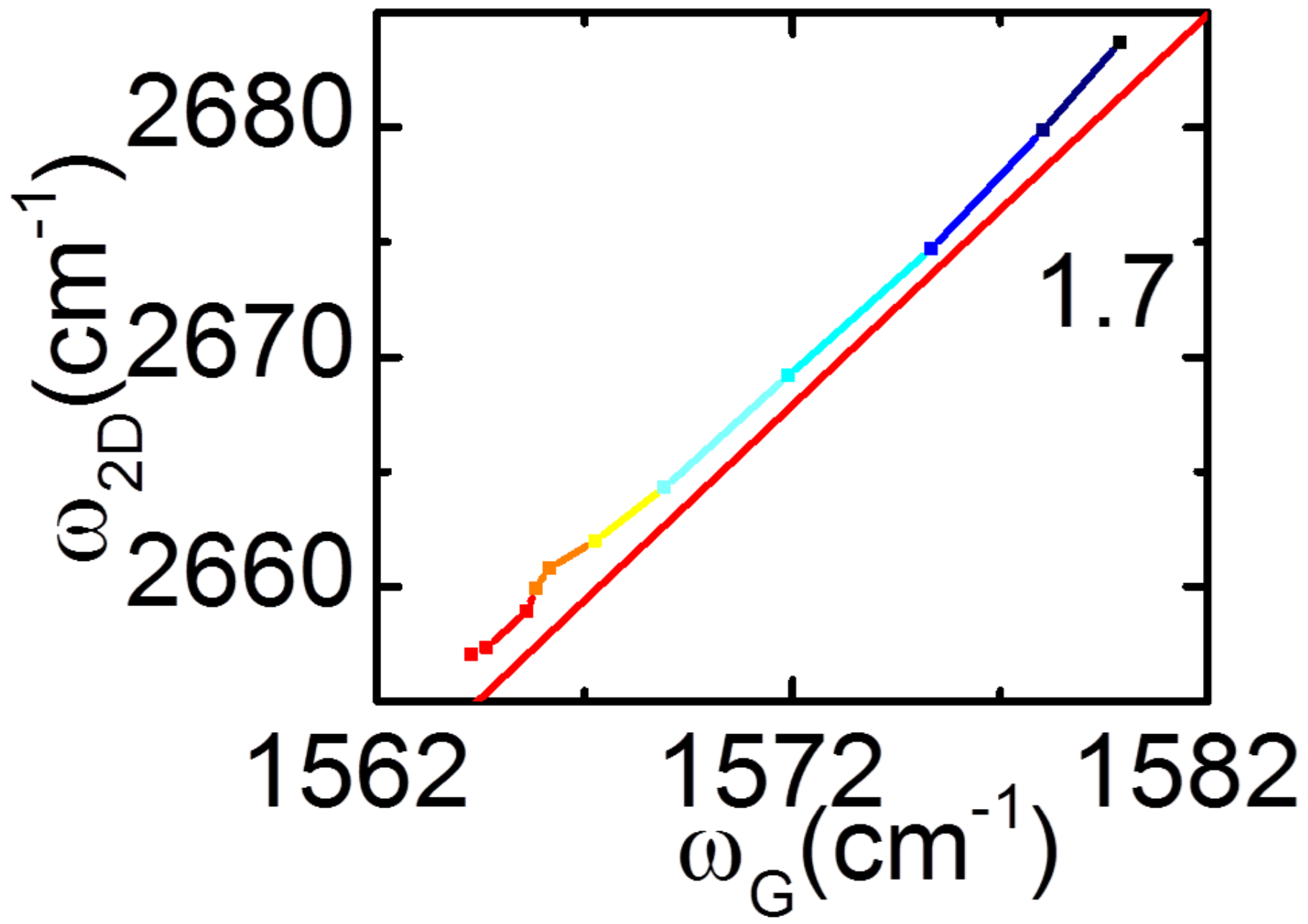

**Figure S4c:** Raman 2D peak position in function of G peak position (Lee et al. diagram)[14] for a full membrane of graphene similar to the figure 4d (without noticeable G peak shift or splitting at low laser power) for different laser power. We can see it fit very well with a slope R=1.7 like expected from a heating variation.

**Number of layers:** The number of layer was determined through the width of the 2D peak, the relative amplitude of the 2D over G peak and the global form of the 2D peak. It stays a quite difficult task. Good references, for comparison are ref [15,16] and typically in most of our sample the thickness was determined to be varying between 2 or 3 layers. In figure S4b there was 2 layer in zone A and 1 in zone B due to the 2D peak form (it has to be note that here, with this type of graphene layer removal, the difference is not necessarily 1) and in figure S9 and S11, for large scale mapping, the number of layer vary between 2 and 3 (here the difference is necessarily 1 layer). This last comment is in accordance with previous samples growth and the n-n+1 step proper to epitaxial growth of graphene.

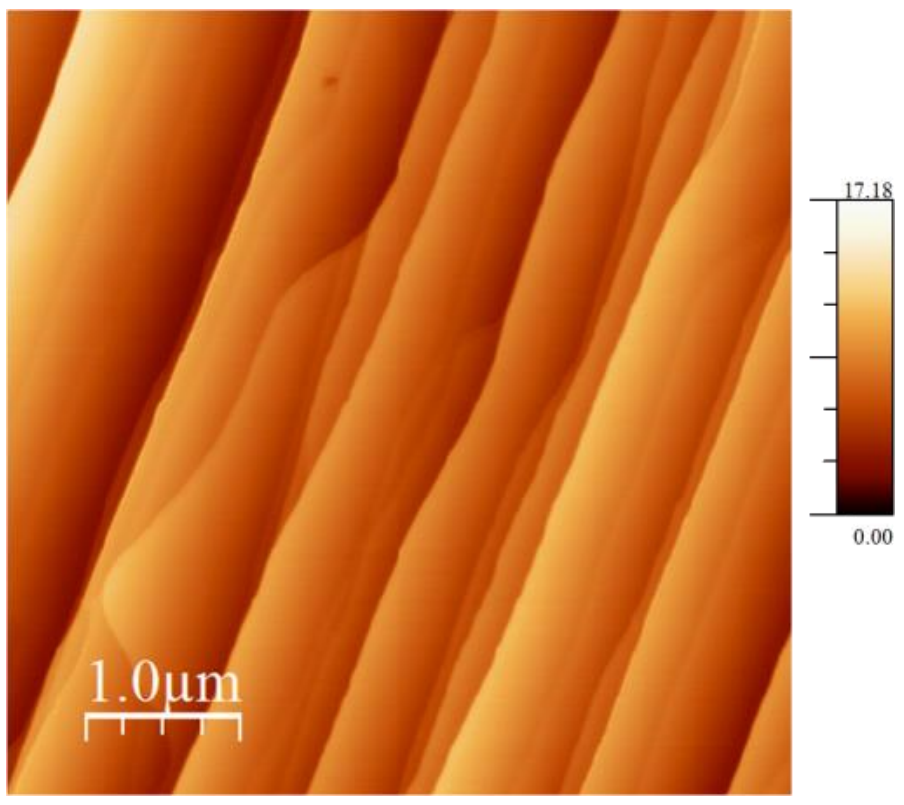

**Figure S4d:** AFM image of epitaxial graphene before the PEC etching. We observe terraces with flat and homogenous graphene

**Quantity of samples:** It is possible to accumulate the number of devices in order to improve the geometry by a simple statistic. For example the structure with the nanoconstriction (figure 3) was obtained with a success of 10% (considering narrower constriction), but this number was not even a limitation in our fabrication process considering the large number of similar structures we have done (>30). Standard membrane were bar typically with 6 µm for the length and 3 for the width, the success rate to obtain these structure is above 90%. Different opening of holes and rectangle were executed in the structure. We have usually 1500 patterns per millimeter square. Long structures reach 200µm by side for square or ribbons (see figure S1).

## S5: Raman difference between doping, strain and temperature dependence

We have to determine a methodology in order to discriminate Raman peak shift due to the different aspects as strain, doping, temperature, number of layers… We will use the table S5 in order to summary every different aspects.

| | $\Delta\omega_{2D}/\Delta\omega_G$ | Splitting peak G | Geometry dependance | Polar dependance | G peak dependance |
|---|---|---|---|---|---|
| Strain | 2.5-3.5 | Yes, for uniaxial strain | yes | yes | 11-32 $cm^{-1}$/%, |
| Doping | < 0.7 | Yes, only for n=2.p layers | at the edges | no | 10 $cm^{-1}/10^{13}cm^{-2}$ |
| Temperature | ~1.7 | No | yes | no | -0.015$cm^{-1}$/K |

**Figure S5:** Dependence of strain, doping or temperature on the Raman G peak shift with also the ratio between the 2D and the G peak, and the G peak splitting behavior [14,17,18,16,19–23] .

**Peak Shift variation**: In this work, we have defined a new method which consists to compare the shift of the G and the 2D peak for each sample in function of the spatial position. In most of our devices, the ratio between these two shifts is between 2 and 3 by varying the position (figure 2 and 4) which directly implies a main contribution from the strain or the temperature. This ration doesn't surpass 0.7 for doping behavior, and can be even negative for large n doping variation [17]. Also, strong shift of the Raman peaks are difficult to explain with doping alone because it imply doping change of more than $10^{13}/cm^{-2}$.

**G peak splitting:** Another pattern on Raman measurement which clearly indicates the strain to be the main contribution in our devices is the splitting of the G peak (or the width of the main G peak, for small variation) which is clearly observed in most of our devices. This effect has been mainly observed under uniaxial strain variations [19,20] and does not appear for biaxial strain[24]. Some splitting of the G peak was also observed due to doping effect, due to peculiar properties of the electron phonons coupling and the whole symmetry of the system. The bilayer exhibits inversion-symmetry breaking because of differential doping between the layers. A graphene with an even number of layers can present a splitting under strong doping superior at less to $10^{13}cm^{-2}$ [18,25,26]. It has to be noticed that the scenario of doping-induced-Raman-resonance-splitting was the chosen one for the authors of ref [2] to explain their G peak splitting with samples following the same technic of fabrication. We don't exclude this possibility to exist in our sample, we have also assume a doping difference between layer in our simulations (see S9) but we observe strong variation of this splitting (>10cm-1) with position around our nanoconstriction. This doping-induced-splitting scenario is not reasonable to explain the entire G peak splitting because strong doping variation on the bottom layer superior at less to $10^{13}cm^{-2}$ have to be assumed in a reproducible way around different nano-constrictions or with sample in fig 3, with splitting modulation up to 20cm-1. Finally, about the temperature dependence, no G peak splitting has been observed for the moment under specific heating of the graphene.

We can do some remarks concerning our devices;
- We identified clearly the pic at 1620 $cm^{-1}$ (fig. 2) and no confusion was done with the G peak splitting and this peak
-The number of layer in epitaxial graphene varies from n to n+1, and the splitting of the G peak is not strongly affected by the number of layer in our measurements. For large samples, some variation appears with position but no real extinction related to the number of layers. For small samples, with, statistically, we must have samples with n layers and some samples with n+1 layers on the whole sample surface, our statistic have shown some reproducible behavior which are not affected by the number of layer concerning the G peak splitting.
-The spatial distribution of the G peak splitting amplitude corresponds quite well with our simulation of strain in our devices, especially for the case of high strain devices with the nanoconstriction. In the case of doping splitting, this spatial distribution of the G peak splitting is not correlated to any specific geometry.
-The quality of our devices (small D peak) and the value of the position for the Raman peak (around $1580cm^{-1}$), under no stress, which is near the neutral point of the G peak indicate both a small initial doping and no oxidation or other defective chemistry.
- We have seen this splitting also in the width increase of the 2D' peak around a graphene nanoconstriction (fig S11) and this peak is said to be much less sensitive to doping effect [27] than the G peak.

Both analyses of the ratio between the peak shift and the G peak splitting and a quantitative and spatial correspondence with strain simulation confirm independently the strain dominance on our measurement concerning the Raman shift we observe in our samples.

**Stress simulation:** At the end, most of our measures correspond also quantitatively to our COMSOL simulation of strain in our devices, if one considers the native compressive stress in epitaxial graphene [7,28] of -2.27 GPa and the different geometry we have used here (see S9).

## S6: strain release

Epitaxial graphene before SiC etching is naturally under compression with a native stress of -2.27GPa [7,28] due to the substrate and graphene lattice period mismatching. In order to control the strain in our structure, the first step is to demonstrate the capability of releasing this internal stress by removing the substrate. By nanostructuring the graphene, it was possible to obtain quasi free standing graphene in air or cantilever like in figure 1ed, S6a or S6b. With simple geometrical consideration, it is possible to relate the residual layer strain ε and the radius of curvature of our structures R trough the interlayer thickness t. Even for highly released graphene with high curvature (fig 1d), the resulting relative elongation of the external layer is very small ε=t/R=3e-4 (see S6a).

**Internal stress and bending contribution** In order to relate the strain and residual stress, it is

$$\Delta T = \frac{E {t_{top}}^2}{6.\, t_{bottom}(1-\nu)R}$$

possible to assimilate the sample to a stress thin membrane on a rigid substrate, where the relative interlayer shears stress ΔT is defined by the Stoney equation:
With E the young modulus=1TPa and ν the Poisson ratio=0.17, it gives a stress of the order of 100kPa much less than the initial stress of -2.27GPa. With our system we begin to deviate from the Stoney equation usual range of applications but it has working quite well for similar AFM plate cantilevers [29]. A more detailed analyze consist to atomistically simulate the intrinsic graphene flake elongation when submitted to an interlayer shear stress [30]. If we consider the shear modulus to be 4.6G Pa in ref[30] , the internal stress we can estimate is around 1MPa. On can be suspicious about the real value of the shear modulus in our devices if we consider the discrepancy between expected spring constant or bending stiffness and measures one for large flake of multi domains CVD graphene[31] but interlayer vibration and mechanic is usually well characterized by Raman spectroscopy (for example in MoS2 or with the C peak in graphene) even in few layers materials and match very well with the expected theoretical values. In the case we consider there is still some SiC under the graphene (in order to exaggerate, we can take 10nm), the residual stress is still equivalent. It doesn't affect the result. $E_{SiC}$ is around 400GPa, $t_{bottom}$ becomes 10nm and $t_{top}$ is now 2x0.33nm.

This approach is equivalent to previous analytical analyze and it is still a demonstrator of our capability to release the stress inside our sample by at less 3 order of magnitude with a simple nanostructuration.

**External stress and bending contributions**. It has to be noticed that we took care to choose especially this sample for our example, with a high curvature oriented to the top in order to be sure to eliminate any external interactions. In fact, most of our graphene flakes have shown much less curvature and in general oriented to the bottom of the sample (see fig. S6b). In some case, the cantilever seems to simply fall down in the direction of the substrate. The bottom layer is the more constrain initially by the substrate and it seems natural to statistically reproduce this state in the suspended case, after etching and release, in case there is not a total release. Others external interactions with graphene are far away from describing theses curvature: in the naive case of gravity, the force apply to a 1µm square trilayer graphene is of the order of $10^{-17}$N/µm$^2$, for our sample a typical bending of the cantilever is around z=1µm (fig. S6b), and the spring constant associated to this force is k=F/z=$10^{-11}$N/m which is out of the range of reasonable values $10^0$-$10^{-5}$N/m [31]. If we consider an electrostatic interaction through a capacitive coupling, in the case of a bad Ohmic contact between the graphene and the substrate, with a substrate at d=8µm from the graphene, the electrostatic capacitance is $C=\varepsilon_0.S/d \sim 10^{-18}$ F. The electrostatic force is $F=dC/dz.\Delta V^2$. At first approximation in z, dC/dz $=-\varepsilon_0.S/2d^2 \sim -10^{-13}$ F/m. A deflection of 1µm and a potential difference ΔV of 1V correspond to a spring constant $k\sim 10^{-7}$ N/m which could be eventually reasonable if the potential difference was not too unrealistic considering the kOhms contact resistance obtain usually between graphene and Cr/au metallization in the group and typical graphene devices for electronic[32].

**Figure S6a:** (On the left) A MEB image of a quasi-free standing graphene cantilever with a tiny anchoring arm, the central part is a square of 3 by 3µm side and the anchoring is 500nm length and 140nm width. The radius of curvature of the plate is around 3µm directly estimated with this image. Scale bar is 5µm. (On the right) A schematic of a multilayer graphene (two) with a certain radius curvature R. The internal and external layers are submitting to an opposite interlayer shear stress T which induces a relative elongation ε of the external layer and a curvature. Simple geometrical consideration relates R (3µm) with ε and t the interlayer thickness (0.34nm), in this sample ε=t/R=3e-4.

**Figure S6b:** MEB images of different samples with cantilever part which bend to the bottom. In all these part anchored by only one side (on the left and right of the main graphene bar), we can observe a completely released graphene bar with a curvature oriented mostly to the bottom of the sample.

## S7: Strain modulation with bar, with holes: additional datas and informations

In the last examples of figure 4c and 4d, we see modulation with complexes features. It implies a strong G peak splitting coming from anisotropic planar strain ($\varepsilon_A \neq 0$) [19,20,24] or asymmetric doping [18,26,25,2] (for 2n layers graphene) or asymmetric strain between layers. We exclude an additional molecular layer interaction with our graphene [33] or a misinterpretation with the defect D' peak around 1620 $cm^{-1}$.
Fig. 4c represents a scan along the rectangle membrane y-axis (another sample in fig S7c and x axis scans in S7d). It is similar to 4b with two larger bars on each side and one thinner in the middle of width W. A large splitting is observed only in the middle of the two full part for five sample, with W from W/2 to 2W (see figS7c) and it seems our example is the most favorable for a high G peak splitting. Naively, in case of strain, initial compressions on the membrane apply by the two fix anchoring will be concentrate in the two big side arms and the central bar must be released. The splitting emerges at the exact point where a moment must be apply on the membrane. This scenario is complicated by the asymmetry of strain between the two layers.
In figure 4d, we present a quite unique sample of a full membrane where strong modulations of the G peak splitting and peak shift are observed. This sample is almost quite rare. 99% of our full membrane has no modulation and no splitting at all. This type of modulation has been seen in a wrinkled graphene with the formation of an orthogonal buckling wave [34]. It means we have created an artificial strain on this sample due certainly to a small and rare movement of the contact anchoring. A transverse scan (x axis) over the membrane of graphene is shown with fitted 2D and G peak positions. We can observe a clear modulation and correlation of the G peak splitting into G+ and G- and 2D peak position along x with two extreme positions in red and black. The relation between $\Delta\omega_G=\omega_{G+} - \omega_{G-}$ and $\omega_{2D}$ is linear (see S7a) with a slope of $\Delta\omega_G = -1.5\ \omega_{2D}+4050$. We also measure a ratio R above 2 (see S7a), a high G peak splitting (20$cm^{-1}$) (see fig 4), a high 2D peak shift (above 7$cm^{-1}$).

For both sample, doping alone doesn't explain the result since it must represent a huge doping asymmetry of few $10^{13}cm^{-1}$, with a full covering by even number of layers, and it is related to the geometry itself. In fig 4c, the G-, G+ peaks width are 18 and 14 $cm^{-1}$ around 1974cm-1 and 1594cm-1 (at the red position), and the G peak width is 19cm-1 at 1582cm-1 for the black position, respectively 2691cm-1 and 2697cm-1 for the 2D peak position, and area ratio 2D over G of 4 and 3 respectively, this is all in line with an undoped graphene (at less <$10^{-13}$cm-1) [15,17,35].
Doping doesn't explain, in figure 4c, the correlation between the peak splitting centered in the full part of the membrane and the width W of the bars. Like for the sample in figure 4c, the data are incompatible with doping description. The average position of G+ and G- does not shift, on the contrary to the 2D peak and there ratio is clearly above 2 and the ratio of intensity between the 2D and G peak is not correlated to the 2D peak position and stay quite constant. And this measurement has been seen in few samples in the same line and few in other lines which reduce strongly the possibility to have a n+1 layer appearing by coincidence at the bar position for each sample. And a strong doping appearing at the graphene edges (by e beam exposure or PEC technic for example) would affect the G peak splitting and 2D peak position in an inverted situation; no splitting on large area and strong doping in nanoconstriction.
In figure 4d, the spatial modulation of the splitting. Strain simulation of a monolayer membrane without buckling, didn't match completely with the measured datas. In case of full buckling, the stress energy must be released and the strain-induced peak shift and splitting must disappear. It is the reason why we understand our result in a complex mixing of two effects: 1) asymmetrical strain as in fig 4b explanation: a small buckling of the strained bottom layer induce a curvature of the stick top layer and an opposite strain 2) strain matrices in each layer present a general planar anisotropy for each point where orthogonal strain can be of opposite sign.

**Figure S7a:** Additional datas on sample in figure 4d with the optical image of the sample, delimited by the dash line rectangle with the scan along the white arrow. We can see the sample is transparent without SiC. Below it is a Lee et al. diagram with ω2D in function of ωG+ and ωG- in the case of a 2 peak fitting. The blue line represents the mean position of ωG and the blue dash line represents a slope R of 2.2. We have also plot the ω2D=f-ωG+- ωG+) function on the right which can correspond to a linear variation with a slope of -1.5.

**Figure S7b:** A detail of the sample present in figure 4a whit periodic array of holes in a suspended graphene with a measurement of the strain modulation along the sample. (On the upper left), a MEB image of the sample, holes are 400nm of diameter and the array has a period in x and y of 700nm. . **(On the bottom left),** qualitative comparison between experimental value of the G peak intensity and simulated value of the 2D Gaussian laser spot size convolution with the sample structure, this determine the spot size in an appropriate way to be around 700nm. It is still possible to see details like the array of holes and the four anchoring arms with a respective intensity ratio. **(On the upper middle),** it is 2D Comsol simulation of the strain along the membrane for an initial stress of -2.27 GPa in x and y in the whole structure (before the released). We can observe a repartition of the stain along the structure and especially the appearance of positive and negative strain in a frame corresponding to the holes patterns and a difference in the anchoring arms regions. **(On the bottom middle**), the measurement of the G peak shift along the device, concatenated into clusters due to low integration time and some noise in the measurement (red=1578$cm^{-1}$, blue=1576.6$cm^{-1}$, green=1576.11$cm^{-1}$, cyan=1581.5$cm^{-1}$). It corresponds quite well to the strain simulation after considering the Gaussian laser spot size convolution and the sample

frame with anchoring arm s and holes patterns. It is an indication that we measure effectively some strain inside our membrane. It is also confirm with the almost no power dependence and thermal effect here, at low power, of this shift, see figure 4c and edges or doping effect, considering the ratio of 2 between the G pix shift and the 2D peak shift, see figure 2c. The Raman D peak was almost not seen in this device. Finally in the **middle part**, we report the same data than in figure 2c concerning the G peak shift at different positions, from 1 to 8, along the holes patterns and the simulated strain expectation after Gaussian convolution with the laser spot size (initial tension of -2.27 GPa and Raman G peak position under no stress at 1582.5cm$^{-1}$). Datas were accumulated with longer times for noise reduction and over a 2 by 2 pixels matrix (a pixel is 100nm by side). The measures values correspond very well to the simulated values of the strain inside our membranes. **(On the upper right)** Measures data of the 2D peak intensity**. (On the middle right)** simulated data for strain after a convolution with a Gaussian spot size of 700nm**. (On the bottom right)** measurement and spatial distribution of the 2D peak shift after a cluster analyze.

**Figure S7c:** An additional graphene device, on the same line that the one in figure 4c, with a nanobar in the middle. On the left, there is an MEB image (scale bar 1µm) of the devices and on the right a plot of the Raman G peak spectrum intensity in function of the wavenumber and the position along the y axe and the white square in the left part. White round and triangle represent the position of the G+ and G- peak obtained after a fit of the spectrum with two separated Lorentzian. The two insets correspond to a cut along the red and black dot line in the 2D plot; we can observe a splitting of the G peak at the full part of the graphene and no splitting of the G peak at the position of the nanobar. **(on the right)** 2D peak shift measurement along the same axe and the intensity ratio between G over D peaks and G over 2D pics (we define the intensity of G peak to be the strict addition of the intensity of G+ and G-). A defect D peak appeared naturally over the thin bar part due to edges contribution.

**Figure S7d:** A graph which corresponds to datas in fig 4c with the $\omega$G peak position along the x axis for two different sample (in black the sample in fig 4c and in red another sample). For clarity we have also design the position of the 3 bars along the scan in black.

## References SOM